\documentclass[11pt]{article}
\usepackage[utf8]{inputenc}
\usepackage{listings}
\usepackage{color}
\usepackage{graphicx}
\usepackage{longtable}
\usepackage{float}
\usepackage{amsmath}
\usepackage{amssymb}
\usepackage{authblk}
\date{\today}

\begin{document}

% Org-mode is exporting headings to 3 levels.
\title{Exact propagation without analytical solutions  }
\author{Thorsten Pr\"ustel} 
\author{Martin Meier-Schellersheim} 
\affil{Laboratory of Systems Biology\\National Institute of Allergy and Infectious Diseases\\National Institutes of Health}
\maketitle
\let\oldthefootnote\thefootnote 
\renewcommand{\thefootnote}{\fnsymbol{footnote}} 
\footnotetext[1]{Email: prustelt@niaid.nih.gov, mms@niaid.nih.gov} 
\let\thefootnote\oldthefootnote
\begin{abstract}
We present a simulation algorithm that accurately propagates a molecule pair using large time steps without the need to invoke the full exact analytical solutions of the Smoluchowski diffusion equation. Because the proposed method only uses uniform and Gaussian random numbers, it allows for position updates that are two to three orders of magnitude faster than those of a corresponding scheme based on full solutions, while mantaining the same degree of accuracy. Neither simplifying nor ad hoc assumptions that are foreign to the underlying Smoluchowski theory are employed, instead, the algorithm faithfully incorporates the individual elements of the theoretical model. The method is flexible and applicable in 1, 2 and 3 dimensions, suggesting that it may find broad usage in various stochastic simulation algorithms. We demonstrate the algorithm for the case of a non-reactive, irreversible and reversible reacting molecule pair.
\end{abstract}
\section{Introduction}
\label{sec-1}

Naive Brownian dynamics (BD) simulations notoriously suffer on only inefficently modeling a molecule's diffusive motion near boundaries. In fact, tiny time steps are in general necessary to resolve the diffusive behavior close to boundaries with acceptable accuracy, rendering the naive application of the BD scheme highly incapable of accounting for reaction-diffusion systems over biologically relevant time scales. Many particle-based stochastic simulation algorithms seek remedy by employing exact analytical solutions of the Smoluchowski diffusion equation that incorporate suitable boundary conditions (BC) \cite{Edelstein:1993, BarSchieber:2002BS, vanZon:2005p340, vanZon:2005p401, Opplestrup:2006p235, Oppelstrup:2009p144, Prustel:Sim2011, Hummer}. Because the analytical solutions take into account the correct boundary behavior, much larger time steps can be used without sacrifying accuracy. 

Schematically, a particle-based algorithm describing bimolecular reactions consists of three major parts. First, it has to decide if an encounter took place during a simulation time step. Secondly, if there was an encounter, did also a reaction occur? Thirdly, if no reaction occurred, the encounter pair is propagated according to a full solution of the Smoluchowski equation that takes into account the BC. In this general scheme, the step that involves the propagation is the most time consuming one, it involves sampling from a probability density function (PDF) that is given by a series expansion where the individual terms are represented by complicated, numerically difficult to evaluate, integrals. 

This is where the proposed method comes into play: It eliminates the need to sample from a complicated series expansion, without giving up on accuracy or large size of the time step. Importantly, only the Gaussian PDF has to be sampled from, which results in a tremendously faster execution of the propagation move. While other algorithms using Gaussian random numbers to update the particle positions have been suggested \cite{Andrews:2004p84, Hummer}, our approach does not require rescaling of reaction rates based on macroscopic parameters \cite{Andrews:2004p84} and is particularly close to the underlying physics as well as, at the same time, easy to implement. Because the position moves provide \textit{en passant} the first-passage (FP) times, it facilitates the use of small-time approximations \cite{Prustel:Sim2011} of those expressions involving Green's functions (GF) that are employed to decide whether there was a reaction or not. This is important for 2D systems, because in this case even the radial GF are given by complicated integrals that make their numerical evaluation quite inefficient. Finally, although the proposed method is designed to provide a coarse-grained description, its accuracy can easily be enhanced and the method can also be made exact in a natural way.

Regarding reversible reactions, the algorithm includes possible rebindings after dissociation within the same time step. Simulations show that incorporating this effect enhances the accuracy compared to those using only a Poisson model. 
 
The manuscript is structured as follows. First, we give a brief overview of those elements of the Smoluchowski theory \cite{smoluchowski:1917, collins1949diffusion, Goesele:1984, Rice:1985} needed to explain the proposed algorithm. Then, to be specific, we consider a stochastic simulation algorithm that has been described before \cite{Prustel:Sim2011} and that uses exact solutions to propagate a molecule pair. Next, we show how the same can be achieved much swifter but virtually equally accurate. Finally, we consider a few examples and construct, via the proposed simulation method, the PDF corresponding to a non-reactive, irreversible and reversible reacting pair in 3D and 2D, respectively.   
\section{Theory}
\label{sec-2}
We consider an isolated pair of molecules with diffusion constants $D_{A}, D_{B}$ that diffuse around each other. Alternatively, the pair may be described as a point-like particle diffusing with diffusion constant $D=D_{A}+D_{B}$ around a static sphere that may be reactive or non-reactive. To be explicit, we consider the 3D case, but corresponding considerations apply equally well to a molecule pair in 1 and 2D. The PDF that gives the likelihood to find the particle located at $\mathbf{r}=(x,y,z)$, given that it initially was at $\mathbf{r}_{0}$, is the GF $p(\mathbf{r}, t\vert \mathbf{r}_{0})$ of the diffusion equation 
\begin{eqnarray}\label{diffEq}
\frac{\partial}{\partial t}p(\mathbf{r}, t\vert \mathbf{r}_{0}) &=& D \nabla^{2}_{\mathbf{r}} p(\mathbf{r}, t\vert \mathbf{r}_{0}). 
\end{eqnarray}
The GF is subject to the initial 
\begin{equation}\label{initial_bc}
p(\mathbf{r}, t=0\vert \mathbf{r}_{0})=\delta^{(3)}(\mathbf{r}-\mathbf{r}_{0})
\end{equation}
and BC
\begin{equation}\label{inf_bc}
p(r\rightarrow\infty, t\vert \mathbf{r}_{0})=0,
\end{equation}
where $r:=\vert\mathbf{r}\vert:=\sqrt{x^{2}+y^{2}+z^{2}}$. 
The solution to Eqs.~(\ref{diffEq}), (\ref{initial_bc}) and (\ref{inf_bc}) is given by the free-space GF
\begin{equation}\label{free-space-GF}
p_{\text{free}}(\mathbf{r}, t\vert\mathbf{r}_{0})=\tfrac{1}{(4\pi D t)^{3/2}}\,e^{-\frac{(\mathbf{r}-\mathbf{r}_{0})^{2}}{4Dt}}.
\end{equation}
Focusing on a radially-symmetric system, the diffusion equation (Eq.~(\ref{diffEq})) can be written in terms of the radial coordinate $r$ alone
\begin{eqnarray}
\frac{\partial}{\partial t}p(r, t\vert r_{0}, t_{0}) &=& D \bigg[\frac{\partial^{2}}{\partial r^{2}} +\frac{2}{r}\frac{\partial}{\partial r}\bigg] p(r, t\vert r_{0}). 
\end{eqnarray}
Chemical reactions are incorporated into this formalism by BC at the encounter distance. The BC that describe an irreversible and reversible reaction read as follows
\begin{eqnarray}
4\pi a^{2} D \frac{\partial}{\partial r}p_{\text{rad}}(r, t\vert r_{0})\vert_{r=a}&=&\kappa_{a}p_{\text{rad}}(r=a, t\vert r_{0}), \label{RADBC}\\ 
4\pi a^{2} D \frac{\partial}{\partial r}p_{\text{rev}}(r, t\vert r_{0})\vert_{r=a}&=&\kappa_{a}p_{\text{rev}}(r=a, t\vert r_{0}) -\kappa_{d}[1-S_{\text{rev}}(t\vert r_{0})], \quad \label{REVBC}
\end{eqnarray}
and are referred to as radiation \cite{collins1949diffusion, Sano_Tachiya:1979, Rice:1985} and backreaction BC \cite{Goodrich:1954, Agmon:1984, kimShin:1999, TPMMS_2012JCP}, respectively. The radiation BC (Eq.~(\ref{RADBC})) describes an irreversibly, partially reactive boundary. The limiting cases of a completely reactive (absorbing) BC \cite{smoluchowski:1917} corresponds to $\kappa_{a}\rightarrow \infty$, while a non-reactive pair is described by $\kappa_{a}=0$ (reflective boundary). The backreaction BC (Eq.~(\ref{REVBC})) takes into account dissociations also. Clearly, in the limit of a vanishing dissociation constant $\kappa_{d}\rightarrow 0$, the backreaction BC reduces to the radiation BC. The survival probability $S_{\text{rev}}(t\vert r_{0})$ that appears in the backreaction BC is defined by
\begin{equation}
S_{\text{rev}}(t\vert r_{0}) = 4\pi \int^{\infty}_{a}p_{\text{rev}}(r,t\vert r_{0})r^{2}dr.
\end{equation}

The radial GF corresponding to the different BC are denoted by $p_{\text{free}}$, $p_{\text{abs}}$, $p_{\text{ref}}$, $p_{\text{rad}}$ and $p_{\text{rev}}$ and will play a prominent role in the algorithm we will discuss in the following. 
For later use, we provide the expressions for $p_{\text{rad}}$ in 3D \cite{carslaw1986conduction} and $p_{\text{rev}}$ in 2D \cite{TPMMS_2012JCP}
\begin{eqnarray}
&&p_{\text{rad}}(r, t\vert t_{0}) =  \frac{1}{ 8\pi r r_{0}}\frac{1}{\sqrt{\pi D t}}\bigg[\exp\left[-\frac{(r-r_{0})^{2}}{4Dt}\right] + \exp\left[-\frac{(r + r_{0} - 2a)^{2}}{4Dt}\right]  \nonumber\\
& & -\kappa \sqrt{4\pi Dt}\exp\left(\kappa^{2}Dt + (r+r_{0} - 2a)\kappa\right)\text{erfc}\left( \kappa\sqrt{Dt} + \frac{(r+r_{0} - 2a)}{2\sqrt{Dt}}\right)\bigg], \label{radialGF}\quad\quad\\
& & p_{\text{rev}}(r, t\vert t_{0}) = \frac{1}{2\pi}\int^{\infty}_{0}e^{-Dtx^{2}}T(r,x)T(r_{0}, x)x \,dx, \label{radialREVGF}
\end{eqnarray}
where $\kappa:=(\kappa_{a} +4\pi a D)/(4\pi a^{2}D)$ and the function $T(r,x)$ is defined as
\begin{eqnarray}
T(r,x) &=& \frac{J_{0}(rx)\beta(x)-Y_{0}(rx)\alpha(x)}{\sqrt{\alpha^{2}(x) + \beta^{2}(x)}}, \\
\alpha(x) &=& \bigg(x^{2} - \frac{\kappa_{d}}{D}\bigg)J_{1}(xa) + \frac{\kappa_{a}}{2\pi a D}xJ_{0}(xa), \\
\beta(x) &=& \bigg(x^{2} - \frac{\kappa_{d}}{D}\bigg)Y_{1}(xa) + \frac{\kappa_{a}}{2\pi a D}xY_{0}(xa).
\end{eqnarray} 
\subsection{Simulation algorithm}
\label{sec-2-1}
It is important to emphasize that sampling from any of the radial GF does \emph{not} completely determine the updated 3D positions \cite{Hummer}. Rather, to this purpose, one has to use the full solutions describing the angle dependency also. The major drawback is that these full solutions are given by quite unwieldy integrals, which make their evaluation painfully slow. Although, in many situations, the algorithm as a whole is still much more efficient than BD, because large steps can be made, the position update mechanism represents a major bottleneck.
The analytical representation for the full GF describing the diffusion of a point particle around a partially absorbing sphere in 3D is known \cite{carslaw1986conduction} to be
\begin{eqnarray}\label{genericGF3D}
&&p_{\text{rad}}(r, \theta, t \vert r_{0}) = \tfrac{1}{4\pi \sqrt{r r_{0}}}\sum^{\infty}_{n = 0}(2n+1)P_{n}(\cos(\theta)) \times \nonumber\\
&&\int^{\infty}_{0}e^{-Dt x^{2}} F_{n+1/2}(r, x) F_{n+1/2}(r_{0}, x)x\,dx.
\end{eqnarray}
The functions $F_{\nu}$ are defined by
\begin{equation}
F_{\nu}(r, x)=\frac{(2 \tilde{h}  + 1)[J_{\nu}(rx)Y_{\nu}(ax) - Y_{\nu}(rx)J_{\nu}(ax)] - 2 ax [J_{\nu}(rx)Y^{\prime}_{\nu}(ax) - Y_{\nu}(rx)J^{\prime}_{\nu}(ax)]}{\lbrace [ (2 \tilde{h}  + 1)J_{\nu}(ax) - 2 ax J^{\prime}_{\nu}(ax)]^{2} + [(2 \tilde{h} + 1)Y_{\nu}(ax) - 2 ax Y^{\prime}_{\nu}(ax)]^{2}\rbrace^{1/2}}. 
\end{equation}
Here, $\theta$ denotes the angle between the corresponding relative position vectors and one defines
$\tilde{h}:= ha := \kappa_{a}/(4\pi a D)$.
In the following, we will describe an position update method that abandons the use of the full GF altogether. To put that update mechanism into context, we focus on a simulation algorithm described in Ref.~\cite{Prustel:Sim2011} and we briefly summarize its main features.

We consider a molecule, located at $\mathbf{r}_{0} = (x_{0}, y_{0}, z_{0})$ at time $t_{0}$, close to a reactive sphere of radius $a$.
We update the position according to free diffusion, i.e. we sample the Gaussian PDF (Eq.~(\ref{free-space-GF})), which amounts to adding the following increments to the Cartesian coordinates 
\begin{eqnarray}\label{BD_move}
x &=& x_{0} + \sqrt{2D\Delta t}\mathbf{N}(0,1), \\
y &=& y_{0} + \sqrt{2D\Delta t}\mathbf{N}(0,1), \\
z &=& z_{0} + \sqrt{2D\Delta t}\mathbf{N}(0,1),
\end{eqnarray}
where $\mathbf{N}(0,1)$ denotes a random number sampled from a Gaussian with vanishing mean and variance equal to unity. 
After this position update, at time $t_{0} + \Delta t$, the molecule is located at $\mathbf{r} = (x, y, z)$. If $r=\sqrt{x^{2}+y^{2}+z^{2}} \leq a$, an encounter took place during the time step $\Delta t$ with probability one. However, as is well-known, even if $r>a$, there may have been an encounter. To take into account those encounters as well, one may employ radial GF \cite{Prustel:Sim2011}. More precisely, the expression
\begin{eqnarray}\label{bayes1D}
1 - \frac{p_{\text{abs}}(r, \Delta t\vert r_{0})}{p_{\text{free}}(r,  \Delta t\vert r_{0})} =: P_{\text{enc}}(r, r_{0}, \Delta t)
\end{eqnarray}
defines the conditional probability $P_{\text{enc}}$ that there was an encounter during $\Delta t$, given the molecule was propagated according to free diffusion. Thus, to test for an encounter, it is sufficient to sample a uniform random number $\xi$ and to check whether $\xi < P_{\text{enc}}$.
On the other hand, if there was no encounter, nothing else has to be done and the next position update can be executed. However, if an encounter was detected, the position $\mathbf{r}$ has to get corrected (as it corresponds to the free-space GF) and one resamples the position according to the expression 
\begin{equation}\label{propagator}
p_{\text{ref}}(\mathbf{r},\Delta t\vert \mathbf{r}_{0}) - p_{\text{abs}}(\mathbf{r},\Delta t\vert \mathbf{r}_{0}),
\end{equation}
that involves the full GF (Eq.~(\ref{genericGF3D}) with $\kappa_{a} = 0$ and $\kappa_{a} =\infty$, respectively).
The rationale to substract the GF with absorbing BC is that $p_{\text{abs}}(\mathbf{r},\Delta t\vert \mathbf{r}_{0})$ accounts for exactly those particle trajectories that never encountered the boundary during the time step. 
Finally, the algorithm has to decide whether an encounter also led to a reaction. To this end, one employs the conditional reaction probability, given that the molecule was propagated according to Eq.~(\ref{propagator}), defined by
\begin{equation}\label{p_reac}
P_{\text{reac}}(r, r_{0}, \Delta t ) =1-\frac{p_{\text{rad}}(r, \Delta t | r_{0}) - p_{\text{abs}}(r, \Delta t | r_{0})}{p_{\text{ref}}(r, \Delta t | r_{0}) - p_{\text{abs}}(r, \Delta t | r_{0})}.
\end{equation} 
This completes a simulation step. As emphasized before, the by far most time consuming step is the propagation according to Eq.~(\ref{propagator}). 
\subsection{Update method without full Green's function}
\label{sec-2-2}
Now we will show how the time-consuming sampling can be avoided and replaced by an iterative scheme that is quite easy to implement. We begin by recalling that $p_{\text{ref}}(\mathbf{r},\Delta t\vert \mathbf{r}_{0})$ can be numerically approximated by repeatedly sampling from $p_{\text{free}}(\mathbf{r},\Delta t\vert \mathbf{r}_{0})$ until one obtains $r>=a$, i.e. all samplings resulting in $r<a$ are rejected and redrawn \cite{Zhou:1990, Hummer}. This sampling scheme will result in a renormalized free-space probability density 
\begin{equation}
\frac{p_{\text{free}}(\mathbf{r}, \Delta t\vert \mathbf{r}_{0})}{\int^{\infty}_{r > a}p_{\text{free}}(\mathbf{r}, \Delta t\vert \mathbf{r}_{0})d\mathbf{r}}. \nonumber
\end{equation}
Now, as $r_{0}\rightarrow a$ or as $\Delta t\rightarrow 0$, one has
\begin{equation}\label{ref_free}
p_{\text{ref}}(\mathbf{r}, \Delta t\vert \mathbf{r}_{0}) \rightarrow \frac{p_{\text{free}}(\mathbf{r}, \Delta t\vert \mathbf{r}_{0})}{\int^{\infty}_{r > a}p_{\text{free}}(\mathbf{r}, \Delta t\vert \mathbf{r}_{0})d^{3}\mathbf{r}}.
\end{equation}
At first, this seems to indicate that this relation cannot be exploited because, first, we would like to keep large time steps and second, in a simulation one faces in general a situation where $r_{0} > a$ and accurately propagating the system closer to the boundary requires exactly the measures (either small time steps or use of exact full solutions) we seek to avoid. So the central question is how we can make use of Eq.~(\ref{ref_free}) when $r_{0}>a$ and $\Delta t$ is large?

To this end, we recall that the expressions appearing in the definition of the reaction probability (Eq.~(\ref{p_reac})) and in Eq.~(\ref{propagator}) and that therefore play an essential role for propagation and detection of reactions in the algorithm, can be written as a convolution relation where the individual factors allow for a clear physical interpretation \cite{Prustel:qm2015}
\begin{eqnarray}\label{fundamentalRelation}
&&p_{\text{ref, rad}}(r, t\vert r_{0}) - p_{\text{abs}}(r, t\vert r_{0}) = \nonumber\\
&& \pm D^{2}\int^{t}_{0}dT\int^{T}_{0}d\tau\, \frac{\partial p_{\text{abs}}(r, t-T\vert \xi)}{\partial \xi }\bigg\vert_{\xi=a}
p_{\text{ref, rad}}(a, T-\tau\vert a)\frac{\partial p_{\text{abs}}(\xi, \tau\vert x_0)}{\partial \xi }\bigg\vert_{\xi=a}.
\end{eqnarray}
This decomposition relation motivates to search for an algorithm that is capable of faithfully constructing the individual processes represented by the rhs of Eq.~(\ref{fundamentalRelation}), upon sampling from a free-space density functions only.
In case of a reflective boundary, sampling according to the renormalized free-space density corresponds to the first two factors on the rhs. We now detail an iterative bisection method to faithfully model the FP time process described by the third factor on the rhs. A similar construction has been described in \cite{ottinger1996stochastic} to determine the last reflection time for a 1D problem in a different context, but it works equally well in 2 and 3D for our problem at hand.  
 
We reconsider the first part of the previously described algorithm, the detection of encounter events via the conditional probability $P_{\text{enc}}$ (Eq.~(\ref{bayes1D})). This method is not exact \cite{Prustel:Sim2011}, an error remains due to the ignorance about the precise value of the FP time $\tau_{\text{FP}}$, i.e.~the time when $r=a$ for the first time. Put differently, the detection of an encounter event via $P_{\text{enc}}$ solely provides an upper bound, i.e.~we only know that $t_{0} < \tau_{\text{FP}}< t_{0} + \Delta t $. However, it turns out that one can determine the FP time and, in addition, \emph{en passant}, the associated full 3D position $\mathbf{r}_{\tau_{\text{FP}}}$ with any desired accuracy, only upon using $P_{\text{enc}}$ and sampling random numbers from the uniform and Gaussian densities.

To see this, we assume that after a position move, the molecule is located at $\mathbf{r}=(x,y,z)$ at time $t_{0} + \Delta t$. If $r<a$, there was an encounter. As we have discussed before, even if $r>a$, an encounter may have taken place. Hence, we test for encounter by employing $P_{\text{enc}}$ as usual. If there was an encounter, but $r>a$, we map the point $\mathbf{r}$ outside the sphere to a point inside the sphere by defining 
\begin{equation}\label{reflection_map}
\mathbf{r} \rightarrow \frac{a-\epsilon}{a+\epsilon}\mathbf{r}, \quad \epsilon:= r-a.
\end{equation}
Next, we split the time interval $[t_{0}, t_{0}+ \Delta t]$ into $[t_{0}, t_{0}+ \Delta t/2]$ and $[t_{0}+\Delta t/2, t_{0} + \Delta t]$. Then, we employ the conditional PDF that describes the intermediate position $\mathbf{r}_{M}$ at time $t_{0}+\Delta t/2$, given that the molecule was at time $t_{0}$ at $\mathbf{r}_{0}$ and ended up at time $t_{0} + \Delta t$ at $\mathbf{r}$. One can show that this conditional probability density is again a Gaussian with first moment given by $(\mathbf{r}_{0}+\mathbf{r})/2$ and variance $\sigma^{2} \rightarrow \sigma^{2}/2$ as follows \cite{ottinger1996stochastic}. Consider two independent Gaussian random variables $\mathbf{X}$ and $\mathbf{Y}$
with variance $\sigma^{2}$ and vanishing mean. To be definite, we consider the 3D case, but the described construction works equally well in 1 and 2D. The quantity we are interested in is the conditional PDF $p(\mathbf{X}= \mathbf{x} \vert \mathbf{Z}:=\mathbf{X} + \mathbf{Y} = \mathbf{z})$. In the context of a simulation, the variable $\mathbf{Z}$ describes the position update $\mathbf{r}-\mathbf{r}_{0}$ during a time step $\Delta t$, hence $\mathbf{X}$ can be interpreted to describe an intermediate position update $\mathbf{r}_{M}-\mathbf{r}_{0}$ during $\Delta t/2$. The conditional PDF can be obtained via
\begin{equation}
p(\mathbf{x}\vert \mathbf{z}) = \frac{p(\mathbf{x},\mathbf{z})}{p(\mathbf{z})},
\end{equation} 
where $p(\mathbf{x},\mathbf{z})$, $p(\mathbf{z})$ refer to the joint PDF of $\mathbf{X}$ and $\mathbf{Z}$ and the marginal PDF of $\mathbf{Z}$, respectively. Using the identities $\langle\mathbf{X}^{2}\rangle =
\langle\mathbf{X}\cdot\mathbf{Z}\rangle = \sigma^{2}$, $\langle \mathbf{Z}^{2}\rangle=2\sigma^{2}$ and $\langle\mathbf{X}\cdot\mathbf{Y}\rangle = \langle\mathbf{Z}\rangle = 0$, one obtains
\begin{equation}
p(\mathbf{x}\vert \mathbf{z}) = \frac{1}{\sqrt{(2\pi)^{3}}\sigma^{3}/2^{3/2} } \exp{\bigg[-\frac{1}{2}\frac{(\mathbf{x}-\mathbf{z}/2)^{2}}{\sigma^{2}/2}\bigg]}.
\end{equation}
Hence, it turns out that the conditional PDF assumes the form of a Gaussian with mean $\mathbf{z}/2$ and variance $\sigma^{2}/2$.
This means, in the context of the simulation, that an intermediate position can be obtained by a naive BD position move (Eq.~(\ref{BD_move})), with adjusted mean and standard deviation). If the in this way constructed intermediate point is $\mathbf{r}_{M}$ with $r_{M} < a$, the FP time has to lie between $t_{0}$ and $t_{0}+\Delta t/2$. But even if $r_{M} > a$, there may have been an encounter in the interval $[t_{0}, t_{0}+\Delta t/2]$. We test for this as usual by invoking $p_{\text{enc}}$. In case there was an encounter, we map the intermediate point inside the sphere (Eq.~(\ref{reflection_map})) and iterate the procedure for the interval $[t_{0}, t_{0}+\Delta t/2]$ and for $\mathbf{r}_{0}$ and $\mathbf{r}_{M}$. In case there was no encounter in that time interval, we have to conclude that the encounter occurred during $[t_{0}+\Delta t/2, t_{0}+\Delta t]$ instead, and we apply the algorithm to that time interval and, correspondingly, to $\mathbf{r}_{M}$ and $\mathbf{r}$. Thus, iteratively, we can determine the \textit{exact} FP time and the \textit{exact} full 3D position of the molecule when its distance first assumes the value of the encounter radius, only using naive BD moves. Next, being located at the encounter radius, we may apply the free propagator with rejection to construct the reflective PDF, meaning that again we only have to employ sampling from a Gaussian density. Note that for this move we have to use the remaining time step, i.e. $\Delta t - \tau_{FP}$, showing that the first-passage time, that we were initially not interested in, plays now an important role. In fact, if one used $\Delta t$, a significant error would result. Finally, it is determined whether a reaction took place via the reaction probability (Eq.~(\ref{p_reac})). Note, however, that now one may use $P_{\text{reac}}(r, r_{M}, \Delta t - \tau_{\text{FP}})$ (compare also with Eq.~(\ref{fundamentalRelation})) instead of $P_{\text{reac}}(r, r_{0}, \Delta t)$. 
\subsection{Reversible reactions}
After the two molecules assumed a bound state, they may dissociate again in the case of an reversible reaction. The dissociation probability within a time step $\Delta t$ is assumed to be given by
\begin{equation}\label{pDISS}
P_{\text{diss}}(\Delta t) = 1-\exp(-\kappa_{d}\Delta t).
\end{equation}
However, especially when the simulation time step is relatively large, the dissociated molecule may rebind again within $\Delta t$. To correct for these rapid rebindings, one may use Eq.~(\ref{p_reac}) to take into account the possibility for reactions. Thus, the algorithm incorporates dissociations in the following way. First, a uniform random number $\xi$ is sampled and compared with the dissociation probability (Eq.~(\ref{pDISS})). If a dissociation took place, one determines the dissociation time $t_{\text{diss}}$ by solving $\xi = P_{\text{diss}}(t_{\text{diss}})$. Then, the dissociated molecule is placed at $r_{0} = a$. The molecule is propagated as described previously, where the propagation time step is $\Delta t - t_{\text{diss}}$. Finally, one employs $P_{\text{reac}}$ for the time span $\Delta t - t_{\text{diss}}$ to check whether there was a rebinding. 

As we will see in the next section, skipping the test for rebindings, can lead to a large error (Fig.~\ref{fig:pREV}, right panel). 
\section{Simulation results}
\begin{figure}[H]
\centering
\includegraphics[scale=0.35]{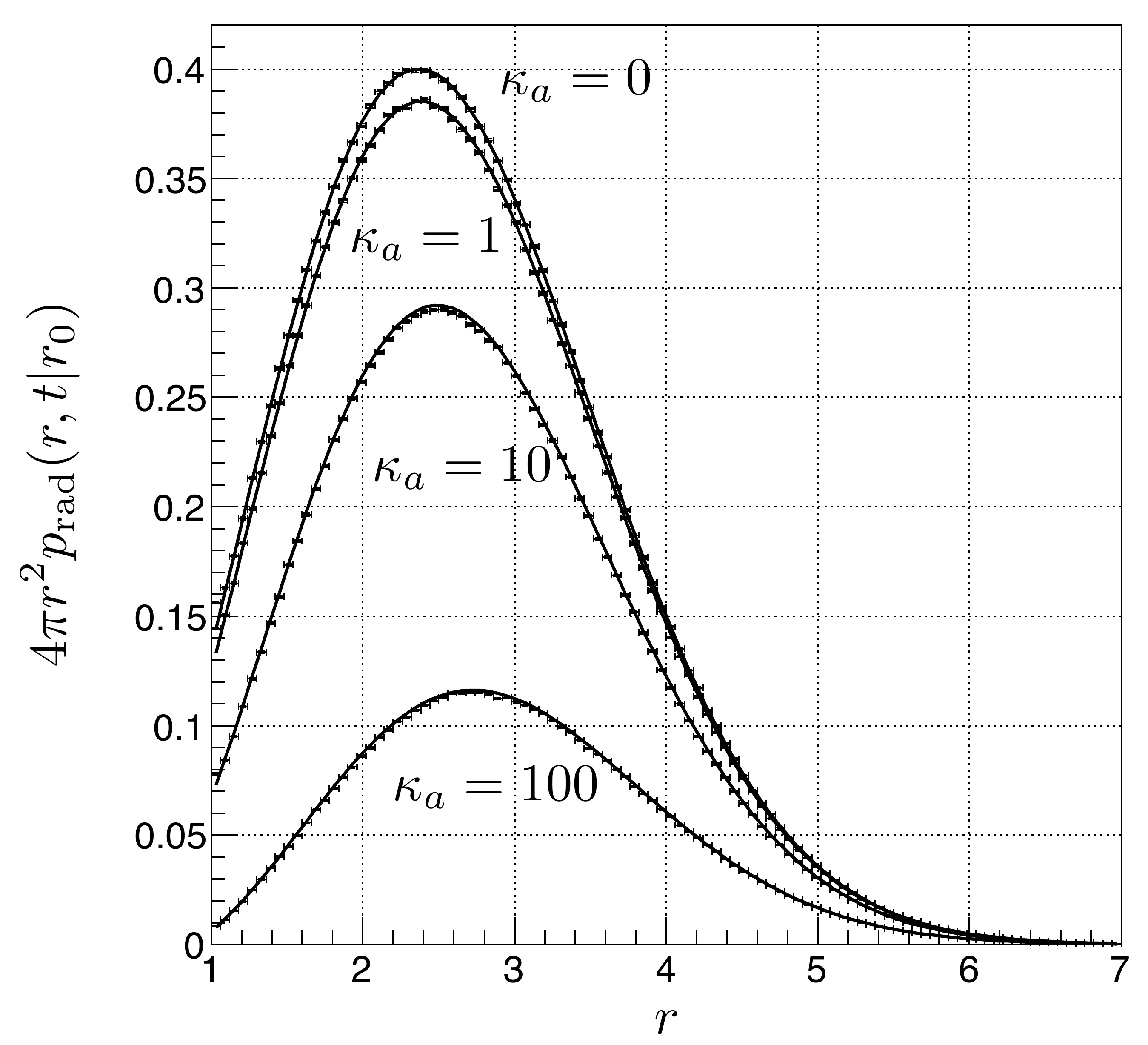}
\caption{Numerical construction of the irreversible PDF $4\pi r^{2} p_{\text{rad}}(r,t \vert r_{0})$ for an isolated pair in 3D. The simulation time and the time step are $t_{sim} = 1$ and $\Delta t = 0.01$, respectively. The other parameters are $D=a=1$, $r_{0}=1.1$. The four curves correspond to different values of the association constant $\kappa_{a} = 0, 1, 10, 100$. The markers indicate the height of the histogram and the solid lines refer to the exact analytical representation of the irreversible 3D GF (Eq.~(\ref{radialGF})).}
\label{fig:pRAD}
\end{figure}

\begin{figure}[H]
\centering
\includegraphics[scale=0.35]{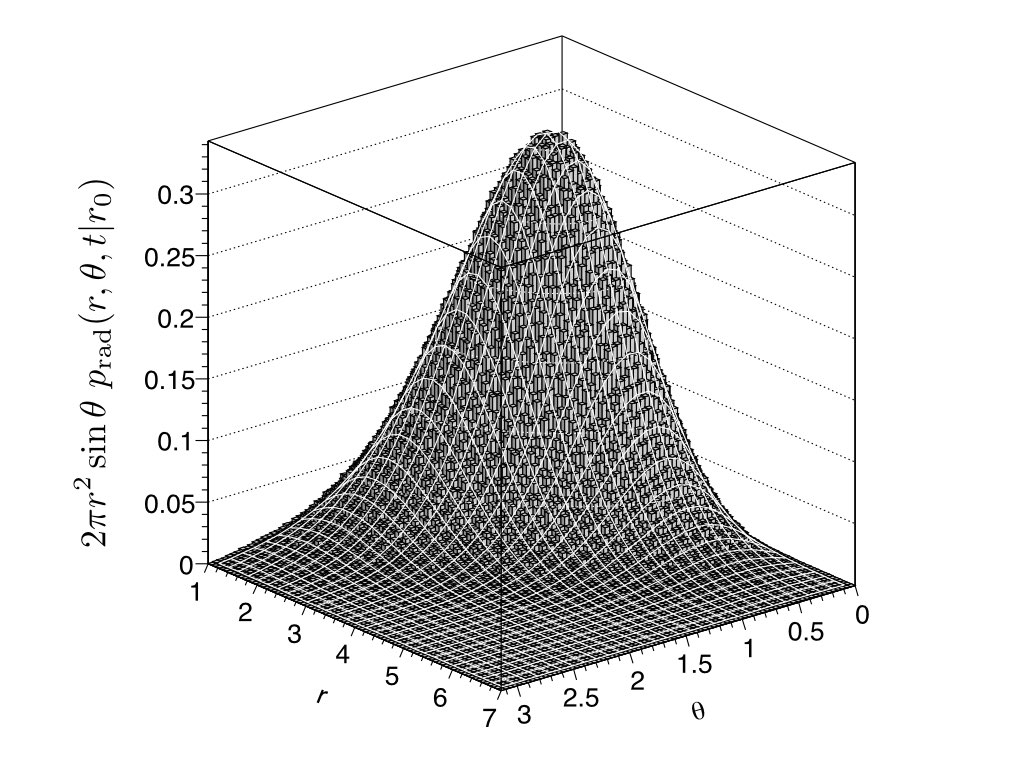}
\caption{Numerical construction of the full GF $2\pi r^{2} \sin\theta \,p_{\text{ref}}(r,\theta, t \vert r_{0})$ for an isolated, non-reactive pair in 3D. The simulation time is $t_{sim} = 1$ and the other parameters are $D=a=1$, $r_{0}=1.1, \Delta t = 0.01$. The histogram bars represent the simulation results and the solid lines refer to the analytical representation of the full 3D GF (Eq.~(\ref{genericGF3D}), with $\kappa_{a}=0$).}
\label{fig:pRADFull}
\end{figure}

\begin{figure}[H]
\centering
\includegraphics[scale=0.2]{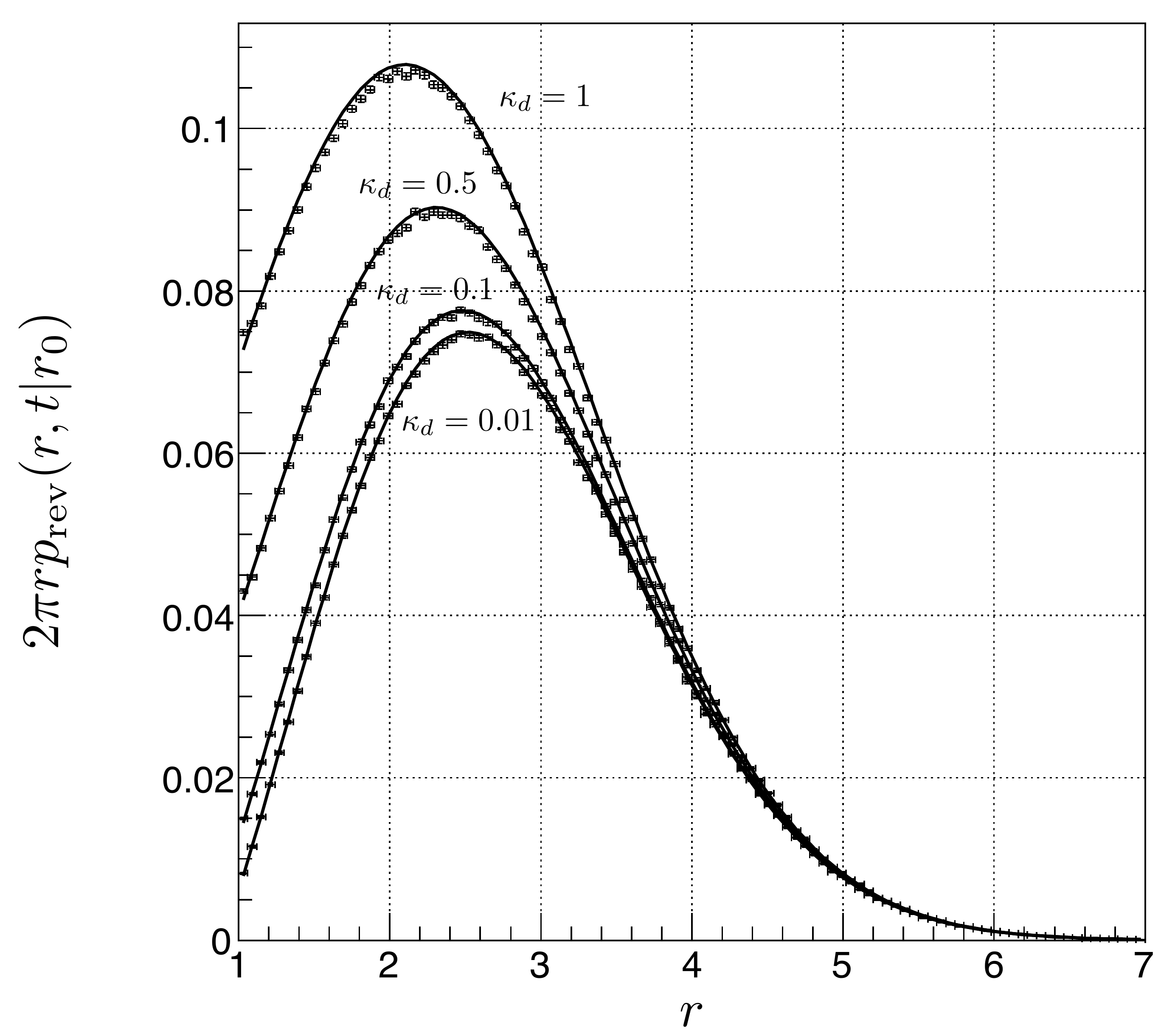}
\includegraphics[scale=0.2]{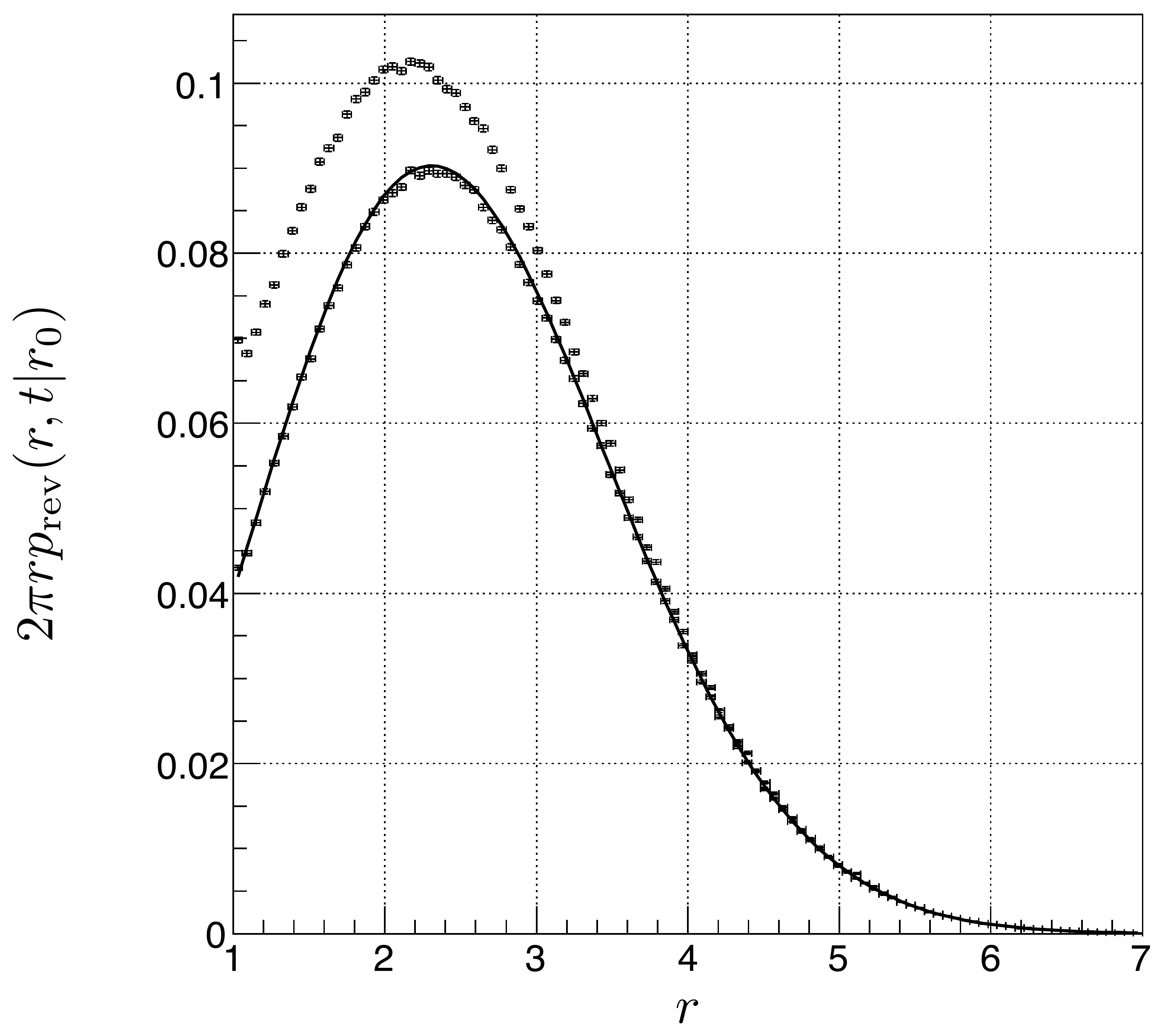}
\caption{Numerical construction of the PDF $2\pi r p_{\text{rev}}(r,t \vert r_{0})$ for an isolated reversible reacting pair in 2D. The simulation time is $t_{sim} = 1$ and the other parameters are $D=a=1$, $r_{0}=1.1, \kappa_{a} = 10, \Delta t = 0.01$. Left panel: The four curves describe the analytical results (Eq.~(\ref{radialREVGF})) for different values of the dissociation constant $\kappa_{d} = 0.01, 0.1, 0.5, 1$. The histogram markers represent the simulation results. Right panel: Simulation results ($\kappa_{d} = 0.5$) obtained by using a Poisson dissociation mechanism only (Eq.~(\ref{pDISS})) show deviations from the analytical curve (Eq.~(\ref{radialREVGF})), while results obtained by additionally including possible rebinding effects show good agreement.}
\label{fig:pREV}
\end{figure}

We employed the described simulation algorithm to construct numerically the GF for the non-reactive and irreversiblly reacting pair in 3D and the reversiblly reacting pair in 2D. 
More precisely, the following simulation set up was used: We considered an isolated pair reacting irreversibly $A+B \rightarrow C$ and reversibly $A+B \leftrightarrow C$. Molecule $A$ was held fixed at the origin. Molecule $B$ was placed at an initial position $\mathbf{r}_{0} = (1.1a, 0, 0)$. The simulation algorithm was run for a time $t_{\text{sim}}$, during which molecule $B$ underwent a diffusive motion described by the diffusion constant $D$. Furthermore, $B$ possibly associated with and dissociated from $A$ according to $\kappa_{a}$ and $\kappa_{d}$, respectively.
After each run, we recorded $B$'s final postion, unless it was bound to A. The corresponding histogram was normalized to account for the number of bound states at $t_{\text{sim}}$. We emphasize that the simulations in 2D (and $\Delta t = 0.01a^{2}/D$) were performed with the small-time expansion of the GF \cite{Prustel:Sim2011}. Fig.~\ref{fig:pRAD} shows the simulation results for the non-reactive and irreversible reacting pair in 3D. The full 3D GF for the non-reactive BC (i.e.~$\kappa_{a} = 0$) is shown in Fig.~\ref{fig:pRADFull}. Finally, Fig.~\ref{fig:pREV} shows the results for the reversible reacting pair in 2D. All simulation results are compared to the corresponding analytical representations (Eqs.~(\ref{radialGF}), (\ref{genericGF3D}), (\ref{radialREVGF})). We find excellent agreement.

\section*{Acknowledgments}
This research was supported by the Intramural Research Program of the NIH, National Institute of Allergy and Infectious Diseases. 
%\bibliographystyle{plain} 
%\bibliography{EG}

\begin{thebibliography}{10}

\bibitem{Agmon:1984}
N.~Agmon.
\newblock {\em J. Chem. Phys.}, 81:2811, 1984.

\bibitem{Andrews:2004p84}
S.S. Andrews and D.~Bray.
\newblock {\em Phys. Biol.}, 1:137, 2004.

\bibitem{BarSchieber:2002BS}
T.M.A.O.M. Barenbrug, E.A.J.F.~(Frank) Peters, and J.D. Schieber.
\newblock {\em J. Chem. Phys.}, 117:9202, 2002.

\bibitem{carslaw1986conduction}
H.S. Carslaw and J.C. Jaeger.
\newblock {\em Conduction of Heat in Solids}.
\newblock Clarendon Press, New York, 1986.

\bibitem{collins1949diffusion}
F.C. Collins and G.E. Kimball.
\newblock {\em J. Colloid Sci.}, 4:425, 1949.

\bibitem{Edelstein:1993}
A.L. Edelstein and N.~Agmon.
\newblock {\em J. Chem. Phys.}, 99:5396, 1993.

\bibitem{Goodrich:1954}
F.C. Goodrich.
\newblock {\em J. Chem. Phys.}, 22:588, 1954.

\bibitem{Goesele:1984}
U.M. G\"osele.
\newblock {\em Prog. React. Kinet.}, 13:63, 1984.

\bibitem{Hummer}
M.E. Johnson and G.~Hummer.
\newblock {\em Phys. Rev. X}, 4:031037, 2014.

\bibitem{kimShin:1999}
H.~Kim and K.J. Shin.
\newblock {\em Phys. Rev. Lett.}, 82:1578, 1999.

\bibitem{Oppelstrup:2009p144}
T.~Oppelstrup, V.V. Bulatov, A.~Donev, M.H. Kalos, G.H. Gilmer, and B.~Sadigh.
\newblock {\em Phys. Rev. E}, 80:066701, 2009.

\bibitem{Opplestrup:2006p235}
T.~Opplestrup, V.V. Bulatov, G.H. Gilmer, M.H. Kalos, and B.~Sadigh.
\newblock {\em Phys. Rev. Lett.}, 97:230602, 2006.

\bibitem{ottinger1996stochastic}
H.C. {\"O}ttinger.
\newblock {\em Stochastic Processes in Polymeric Fluids: Tools and Examples for
  Developing Simulation Algorithms}.
\newblock Springer, Berlin, 1996.

\bibitem{Prustel:Sim2011}
T.~Pr\"{u}stel and M.~Meier-Schellersheim.
\newblock arXiv:1107.0270 [q-bio.QM], 2011.

\bibitem{TPMMS_2012JCP}
T.~Pr\"ustel and M.~Meier-Schellersheim.
\newblock {\em J. Chem. Phys.}, 137:054104, 2012.

\bibitem{Prustel:qm2015}
T.~Pr{\"u}stel and M.~Meier-Schellersheim.
\newblock arXiv:1508.01595 [q-bio.QM], 2015.

\bibitem{Rice:1985}
S.A. Rice.
\newblock {\em Diffusion Limited Reactions}.
\newblock Elsevier, New York, 1985.

\bibitem{Sano_Tachiya:1979}
H.~Sano and M.~Tachiya.
\newblock {\em J. Chem. Phys.}, 71:1276, 1979.

\bibitem{vanZon:2005p340}
J.S. van Zon and P.R. ten Wolde.
\newblock {\em Phys. Rev. Lett.}, 94:128103, 2005.

\bibitem{vanZon:2005p401}
J.S. van Zon and P.R. ten Wolde.
\newblock {\em J. Chem. Phys.}, 123:234910, 2005.

\bibitem{smoluchowski:1917}
M.~von Smoluchowski.
\newblock {\em Z. Phys. Chem.}, 92:129, 1917.

\bibitem{Zhou:1990}
H-X. Zhou.
\newblock {\em J. Phys. Chem.}, 94:8794, 1990.

\end{thebibliography}

\end{document}